\documentclass[12pt,preprint,letterpaper]{aastex}
\usepackage{natbib}

\slugcomment{Accpeted by The Astronomical Journal}

\newcommand{\skipthis}[1]{}

\begin{document}

\shorttitle{Star Formation in RCW 38}
\shortauthors{DeRose et al.}

\title{
A VLT/NACO Study of Star Formation in the Massive Embedded Cluster RCW 38
{\footnote{Based on observations performed at the European Southern
Observatory Very Large Telescope on Cerro Paranal, Chile, under program ID
70.C-0400(A).}}
}

\author{K.L. DeRose, T.L. Bourke, R.A. Gutermuth and S.J. Wolk}
\affil{Harvard-Smithsonian Center for Astrophysics, 60 Garden Street,
Cambridge, MA 02138}

\author{S.T. Megeath}
\affil{Department of Physics and Astronomy, The University of Toledo, 2801
West Bancroft Street, Toledo, Ohio 43606}

\author{J. Alves}
\affil{Centro Astron\'omico Hispano Alem\'an, C/Jes\'us Durb\'an Rem\'on 2-2,
E-04004 Almer\'{i}a, Spain}

\and

\author{D. N\"urnberger}
\affil{European Southern Observatory, Casilla 19001, Santiago, Chile}

\begin{abstract}

We present the results of high angular resolution adaptive optics (AO)
near-infrared ($JHK_s$) observations of the deeply embedded massive cluster
RCW 38 using NACO on the VLT.  Narrowband AO observations centered at
wavelengths of 1.28 \micron, 2.12 \micron, and 2.17 \micron\ were also
obtained.  The area covered by these observations is about 0.5 pc square,
centered on the O star RCW 38 IRS2.  We use the $JHK_s$ colors to identify
young stars with infrared excess in this region.  Through a detailed
comparison to a nearby control field, we find that most of the 337 stars
detected in all three infrared bands are cluster members ($\sim$317), with
essentially no contamination due to background (likely due to the high
cluster extinction of $A_V \sim 15$) or foreground sources.  Five sources
with 3 band detections have colors suggestive of deeply embedded
protostars, while 53 sources are detected at $K_s$ only; their spatial
distribution with respect to the extinction suggests they are highly
reddened cluster members but their evolutionary status is unclear.
Detectable $K_s$-band excess is found toward $29 \pm 3\%$ of the stars.
For comparison to a similar area of Orion Nebula Cluster observed in the
near-infrared, mass and extinction cuts are applied, and the excess
fractions redetermined.  The resulting excesses are then $25 \pm 5\%$ for
RCW 38, and $42 \pm 8\%$ for Orion.  RCW 38 IRS2 is shown to be a massive
star binary with a projected separation of $\sim500$ AU.  Two regions of
molecular hydrogen emission are revealed through the 2.12 \micron\ imaging.
One of these shows a morphology suggestive of a protostellar jet, and is
clearly associated with a star only detected at $H$ and $K_s$, and
previously identified as a highly obscured X-ray source.  Three spatially
extended cometary-like objects, suggestive of photoevaporating disks, are
identified, but only one is clearly directly influenced by RCW 38 IRS2.
The structure of the inner core of RCW 38 is also characterized and
compared to Orion and other clusters.  A King profile provides a reasonable
fit to the cluster radial density profile and a nearest neighbor distance
analysis shows essentially no sub-clustering.

\end{abstract}


\section{Introduction}

The study of embedded clusters is vital to our understanding of star
formation because most stars have been shown to form in large clusters
with hundreds of stars (Lada $\&$ Lada 2003, Porras et al.\ 2003,
Carpenter 2000 and Allen et al.\ 2007).  Of particular interest
are the largest clusters with thousands of stars because it is in
these clusters where large numbers of low mass stars live in close
proximity to luminous O-stars.  In the Lada $\&$ Lada 2003 tabulation
of clusters within 2~kpc, there are two such larger embedded clusters
with over 1000 stars: the Orion Nebula Cluster (ONC) and RCW 38 (Wolk
et al 2006).  The other clusters within 1 kpc typically contain only a
few hundred stars, making direct comparisons with the ONC not
possible.  It is even likely that our Sun was formed in such a
cluster.  A possible explanation for the abundances of many
radioactive isotopes in our Solar System is the presence of a nearby
supernova during its formation which would have been a more probable
occurrence in a rich cluster like the ONC or RCW 38(Meyer \& Clayton 2000).

Studying embedded star clusters not only yields important information
regarding the formation and evolution of stars, it also yields important
information regarding the formation and evolution of planetary systems
around these stars.  As first studied in the Orion Nebula Cluster (ONC),
proplyds (PROto PLanetarY DiskS) exist as large disks of dust and gas that
provide material for future planet formation.  However, when exposed to
unattenuated ultraviolet (UV) radiation the molecular hydrogen dissociates
causing the outer portions of the disk to photoevaporate (Hollenbach et
al.\ 1994).  Since many young low-mass stars in other regions exist in
close proximity to massive OB stars that output a significant amount of UV
radiation, it is important to determine whether this process seriously
inhibits planet formation around them.  Low-mass stars are important
because they represent the majority of young cluster members, which means
the vast majority of planets may be formed around them.  Thus, low-mass
stars clustered around OB sources are at the same time the most likely to
form planets and the most likely to experience photoevaporation effects.
Knowledge of the disk frequency in a cluster as a function of the distance
from the massive OB stars can constrain the affect of photoevaporation on
disk evolution (Balog et al.\ 2007).  Such knowledge will help to constrain
the conditions that are necessary for those disks to evolve into planets.

The HII region RCW~38, at a distance of 1.7~kpc (Beck, Fischer \&
Smith 1991), contains an embedded cluster centered around a O5 star
IRS2.  RCW~38 is more embedded than the ONC and due to the OB stars in
the center it provides a similar environment for the study of disk
photoevaporation and evolution.  Smith et al.\ (1999) showed that the
winds from IRS2 have created a 0.1 pc bubble that is relatively free
of dust, leaving stars within the bubble directly exposed to UV
radiation.  This bubble is filled with diffuse thermal and synchrotron
X-ray emission (Wolk et al.\ 2002) and the intersection of the central
gas bubble with the surrounding dust cloud appears to be a site of
active star formation (Smith et al.\ 1999).

Wolk et al.\ (2006) observed the central region of RCW~38 at X-ray and
near-infrared wavelengths, identifying a large cluster population of 360
X-ray sources that are likely cluster members over the total field down to
Log[L$_x$]$\sim$30.  Comparing this to the COUP (Chandra Orion Ultradeep
Project) measurements of the ONC discussed in Feigelson et al.\ (2007),
roughly 200 sources of their sample are more luminous than
Log[L$_x$]$\sim$30.  The number of sources in the COUP measurement is on
the same order as Wolk et al.\ (2006).  The somewhat larger numbers in the
latter study may be explained by the larger physical area surveyed relative
to the former.  Since the ONC and RCW~38 have similar populations of young
stars, this shows that both the ONC and RCW~38 are unique large clusters
within 2~kpc.  For the central region of RCW~38, Wolk et al.\ (2006)
determined that 97\% of the $\sim$200 X-ray sources within 200$\arcsec$ of
IRS2 are likely cluster members.  In this region they determine that the
disk fraction is 49\% from near-infrared ($JHK_s$) excess measurements.
The infrared observations of Wolk et al.\ (2006) of the central 2\farcm5 of
the RCW~38 cluster were performed with ISAAC at the Very Large Telescope
(VLT), with resolution insufficient to resolve close sources in the most
crowded region around the O5 star.  We present here the results of our
observations of the central $\sim$1\arcmin\ of RCW~38 with adaptive optics
near-infrared imaging with the NACO system on the VLT, which allow us to
better separate close sources and to separate the source emission from the
nebular background emission, providing a sample that is more easily
compared to the nearer ONC.  In \S2 we describe the observations, in \S3 we
discuss the photometery and completeness limits, and in \S4 we present the
results on infrared excess sources, cluster membership, the central O star,
protostellar candidates, molecular hydrogen emission, proplyd candidates,
and the disk fraction.  In \S5 we analyse these results, discussing cluster
structure and the effect of the O stars on disks around low mass stars.
Our summary is presented in \S6.

\section{Observation and Data Reduction} 
\label{obs}

\subsection{NAOS-CONICA}

High angular resolution near infrared imaging of RCW~38 was performed on
February 20, 2003 and February 23, 2003 using NAOS-CONICA (NACO) at Yepun,
one of the 8.2\,m unit telescopes of the Very Large Telescope (VLT)
installed at the European Southern Observatory's (ESO) site on Cerro
Paranal, Chile.  NAOS (Nasmyth Adaptive Optics System) is an adaptive
optics (AO) system that assists imaging at wavelengths in the range
0.8--5.5\,$\mu$m by directing light from the telescope partially into a
(either optical or infrared) wavefront sensor.  The remaining light is
directed into the CONICA infrared camera that is equipped with a
1024\,$\times$\,1024 pixel Aladdin InSb array detector (Lenzen et\,al.\
2003; Rousset et\,al.\ 2003).

The CONICA S54 setup (pixel scale $\sim$ 0\farcs054; instantaneous
field-of-view $\sim$ 56$\arcsec$\,$\times$\,56$\arcsec$) was used to
take 3 sets of 14 frames each in the broad-band $J$, $H$ and $K_s$
filters, centered at a wavelength of 1.27\,$\mu$m, 1.66\,$\mu$m and
2.18\,$\mu$m, respectively.  In addition, 3 sets of 12 frames each
were taken in narrow-band filters centered at the following
wavelengths: 1.28\,$\mu$m (bandwidth $=$ 0.014\,$\mu$m), 2.12\,$\mu$m
(bandwidth $=$ 0.022\,$\mu$m) and 2.17\,$\mu$m (bandwidth $=$
0.023\,$\mu$m), covering the {\it Pa}\,$\beta$, H$_{2}$ (1--0) S(1),
and {\it Br}\,$\gamma$ lines, respectively.  Detector integration
times (DITs) and NDITs are given in Table \ref{dit}.  The integration time per
frame is DIT $\times$ NDIT.

Wavefront sensing was performed with the IR sensor to close the AO loop on
the bright infrared source RCW~38 IRS\,2 (self-referencing) at RA =
08$^{h}$59$^{m}$05$\farcs$5, DEC = $-$47$^o$30$\arcmin$39$\farcs$4 (J2000),
and SpT $=$ O5V (Frogel \& Persson 1974; Smith et\,al.\ 1999).  The airmass
ranged from 1.15 to 1.30 during the observations.  For both nights,
observing conditions were exceptionally good and stable, with a clear sky
and optical seeing mostly in the range 0\farcs4--0\farcs6.  As measured on
the obtained data sets themselves, the image quality (i.e., the median FWHM
PSF of randomly selected point sources) is found to be better than 0\farcs2
for all six filter settings.

\begin{figure}
\notetoeditor{Please print this figure across one page (2 columns)}
\centering
\includegraphics[width=6in]{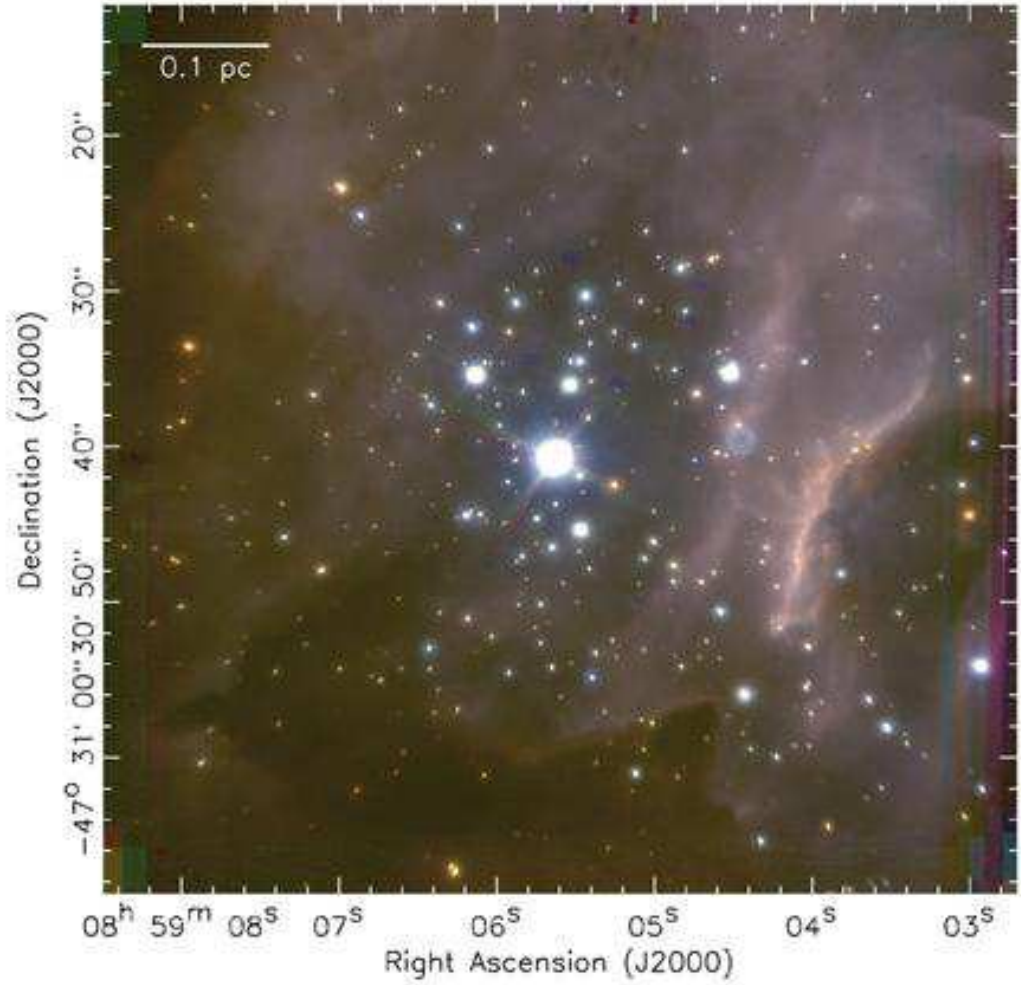}
\caption{Color composite image of RCW~38 through $J$ (blue), $H$ (green), and
$K_s$ (red) filters.  The bright star in the center is IRS2.  The blue sources
to the north and south of IRS2 are J-band artifacts, as is the ring-like
structure to the west of IRS2 in the nebulous filament (known as IRS1 from 
mid-infrared wavelengths).
}
\label{im}
\end{figure}

To allow corrections for cosmic rays and bad pixels, a jitter pattern
(random offsets within a jitter box of width 2$\arcsec$) was applied,
resulting in a total field of view of $60\farcs8 \times 58\farcs3$.
This corresponds to linear scales of 0.5\,pc\,$\times$\,0.48\,pc at
the distance of 1.7\,kpc.  Twilight flats, lamp flats and dark frames
were taken through the usual NACO calibration plan.  For each
individual frame, all basic steps of data reduction (i.e.,
flat-fielding, sky subtraction and bad pixel correction) were
performed using standard routines within the IRAF software package.
For an exhaustive description of the processing of crowded field IR
data we refer to N\"urnberger \& Petr-Gotzens (2002).  Figure \ref{im}
shows the final data set as a combined 3-color image.

\subsection{SOFI Control Field}

In order to estimate the degree of background contamination, a control
field (J. Ascenso, private communication) was imaged on June 17, 2005 in
the $J$, $H$, and $K_s$ passbands using the SOFI (Son of ISAAC) on the NTT
(New Technology Telescope) at La Silla, Chile. The resolution of the image
is approximately 0.29 arcsec/pixel and centered at $10^{\rm h}18^{\rm
m}01\fs2$ and $-57\degr15\arcmin31\farcs1$. A total of 6 frames per
passband were combined together, resulting in an integration time of 36 s
per band.  The total FOV is approximately $4\arcmin~\times~4\arcmin$.

\section{Source Identification and Photometry} \label{sourceid}

\subsection{PhotVis Identification}

Point source detection and photometry was accomplished using Gutermuth's
IDL photometry and visualization tool, PhotVis version 1.10 (Gutermuth et
al. 2008).  PhotVis selects point sources from embedded regions containing
significant amounts of nebulosity with a modification to the standard
DAOphot algorithm.   PhotVis is heavily dependent on the FITS procedures
and DAOphot version 1 port supplied by the IDL Astronomy Users Library
(Landsman 1993).

Gaussian distributions were assumed for all sources and sources with a
full-width at half-maximum (FWHM) greater than 2.5 pixels (0\farcs135) were
discarded. The images were also visually inspected to remove false
positives due to image artifacts.  Flux measurements were taken using a 3
pixel (0.162\arcsec) radius and background measurements were determined
using inner and outer sky annuli of 5 pixels (0\farcs27) and 8 pixels
(0\farcs432) respectively.  This procedure was used to perform relative
photometry on both NAOS-CONICA data and the SOFI control field.  Absolute
photometry was not performed on the NACO narrow band images.

\subsection{Completeness and Detection Limits}

Differential completeness limits of 90$\%$ were calculated in all images by
inserting successively fainter fake stars with gaussian profiles and FWHM
equivalent to those in the image and taking 90$\%$ of the false sources
recovered.  For this test, sources with magnitude errors greater than 0.2
were discarded as well as sources that deviated in brightness by $3\sigma$
more than the original (due it being overlayed over a brighter source).
Final values for the completeness limits of the NACO data are $J$=19.01
mag, $H$=18.13 mag, and $K_s$=16.97 mag.  Detection limits (5$\sigma$) in
these bands are $J$=20.3 mag, $H$=19.0 mag, and $K_s$=18.0 mag.  The 90$\%$
completeness limit for the SOFI control field calculated in the same manner
is $K_s$=15.67 mag.  $J$ and $H$ completeness limits were not calculated
because only the $K_s$-band control field was necessary for background
contamination estimates.  The saturation limits for both fields was
approximately 11th magnitude in all 3 passbands.

Utilizing the $J$ and $H$ completeness limits yields a mass detection limit
of 0.12 $M_\sun$ at 1 Myr for an average extinction of $A_V=10$ ($A_K=1.1$;
Rieke and Lebofsky 1985).  The mass detection limit was approximated using
the Baraffe 1 Myr isochrone (Baraffe et al.\ 2002) and the distance modulus
of 11.15 calculated with the assumed distance of 1.7 kpc.  This detection
limit is right above the hydrogen burning limit of 0.08 $M_\sun$, so there
are no brown dwarf candidates in the NACO sample.

\subsection{NAOS-CONICA Calibration}

The NACO WCS and zero-point flux calibrations were performed using larger
FOV images of RCW~38 obtained with ISAAC (Infrared Spectrometer And Array
Camera) on the VLT.  We used the photometry extracted from the ISAAC data
by Ascenso (2008).  Due to the use of PSF fitting and to a better estimate
of the sky contribution on the nebulous background, the Ascenso (2008)
photometry is believed to be more reliable than the aperture photometry of
Wolk et al.\ (2006).  The ISAAC data was registered and calibrated using
sources from the 2MASS Point Source Catalog in the larger ISAAC field.  The
SOFI control field was flux and position calibrated using 2MASS sources at
measurements up to magnitude 14 in $JHK_s$, due to the fact that the SOFI
field is much deeper.  All sources in all images with a measured
uncertainty greater than 0.2 magnitudes (outside of 1$\sigma$) have been
discarded.

In addition, because the AO feedback loop was closed using a central source
for the NACO data, there is a significant amount of off-axis elongation in
many sources in the images.  This caused a linear trend in magnitude
difference between the ISAAC and NACO photometric measurements as a
function of radial distance from IRS2.  We therefore used position
dependent zero points given by the equations:

\begin{eqnarray}
J: \,\, 24.426 - 0.0176 \times r_{\rm off} \nonumber \\ 
H: \,\, 22.828 - 0.0247 \times r_{\rm off}  \\
K_s: \,\, 21.648 - 0.0274 \times r_{\rm off} \nonumber
\end{eqnarray}

where $r_{\rm off}$ is the radial offset from IRS2 in arcseconds.

\section{Results}

\subsection{Color-color and Color-magnitude Diagrams}
\label{culling}

\begin{figure}
\centering
\includegraphics[width=3.5in]{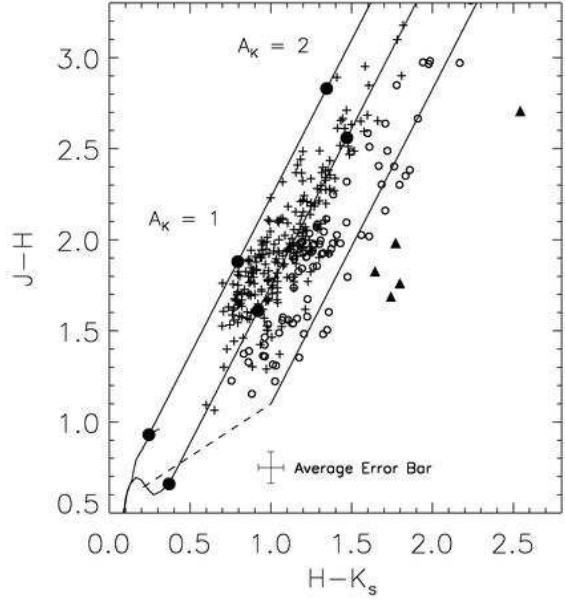}
\caption{Color-Color Diagram for the NACO field.  Plotted are the 337
sources detected in $JHK_s$ as discussed in Section \ref{culling}.  Reddened
sources with $\sigma<0.2$ are displayed with crosses.  Open circles
indicate $K_s$-band excess sources identified by discarding sources within
1$\sigma$ of the lowest reddening vector in either dimension.  Filled
triangles represent protostellar candidates (Class I sources) or extreme
CTTS's (Lada \& Adams 1992 or Carpenter et al. 1997).  Upper and lower
reddening vectors are from Rieke \& Lebofsky (1985) and plotted as solid
lines for an M5 giant (Bessell \& Brett 1988) and M5 dwarf (B.\ Patten
2004, private communication) respectively.  Reddening of $A_K = 1$ and 2 is
indicated by the filled circles.  The classical T-Tauri stars
(CTTS) locus from Meyer et al.\ (1997) is plotted as a dotted line.  Some
of the sources are identified as non-excess sources despite appearing in
quite far into the region where they would appear to be excess sources, due
to greater than average errors that overlap with the non-excess region to
the left of the first reddening vector.  These sources are all faint
sources, and are evenly distributed at different radii from IRS2. 
}
\label{fig-cc}
\end{figure}

\begin{figure}
\centering
\includegraphics[width=3.5in]{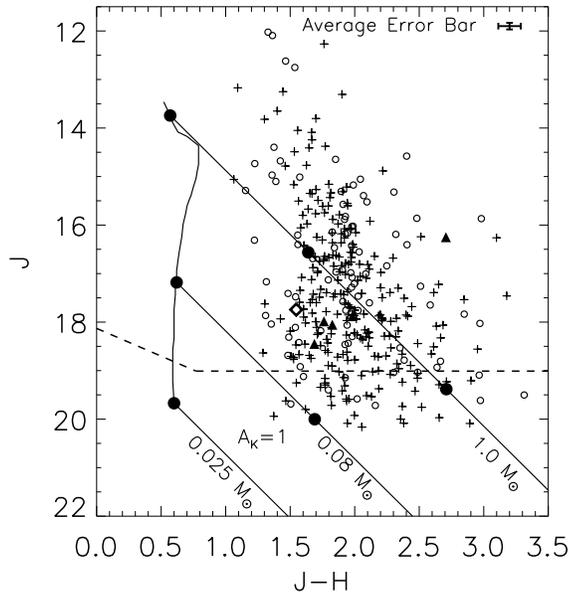}
\caption{Color-Magnitude diagram for NACO field.  Symbols as in Figure
\ref{fig-cc}: reddened $JHK_s$ sources are crosses, $K_s$-band excess
sources are circles, and candidate protostars are triangles.  The single
black diamond indicates the only source detected in $J$ and $H$ that was not
detected in $K_s$ (thus our sample is not biased towards detected excess in
$K_s$).  Completeness limits for $J$ and $H$ are represented as dashed
lines.  The curved line at left is the unreddened 1 Myr pre-main-sequence
isochrone from Baraffe et al. (2002) with reddening vectors (Rieke $\&$
Lebofsky 1985) at 0.025 M$_\sun$, 0.08 M$_\sun$, and 1.0 M$_\sun$ shown as
downward sloping lines.  From this diagram it is clear that the cluster is
reddened by around $A_K=1$.  Despite the surrounding differential reddening
that exists due to the nebulosity, the cluster is surrounded by a bubble of
nebulosity with the central region evacuated due to the wind from the O
stars.  This causes the overall reddening to be fairly constant over the
sources within the cluster core.
}
\label{cmd3}
\end{figure}

The infrared photometry for the 483 sources detected in at least one
band with a magnitude uncertainty of $<$ 0.2 is listed in
Table~\ref{src-list}.  The number of sources detected in each of the
three broad-bands are shown in Table~\ref{sources}.  We detected 409
sources in $H$ and $K_s$ and 344 sources in all bands with
uncertainties $<$ 0.2 magnitudes.  This uncertainty cutoff results in
average photometric errors of 0.07 mag for $J$ and 0.06 mag for $H$
and $K_s$.  Inspection of the 344 sources detected in $JHK_s$ shows
that seven sources have ($J-H$) values less than 1 magnitude and
deviated significantly from the bulk of the sample.  Of these seven,
four have a $K_s$-band excess.  Upon visual inspection, it was evident
that two of these sources had an excess of flux detected in $J$ due
to artifact contamination in that image (where residual images of the
central source IRS2 can be seen in multiple places in the central
portion of the image).  Along the same lines, four of the other
sources were very near IRS2 or partially obscured by a much larger,
brighter star, which explains the excess flux detected in J.  The
final source, that remains unexplained by the previous explanations,
is likely a foreground source (especially due to the fact that it does
not show $K_s$-band excess) and thus, is not a cluster member.  Of
the seven sources, six have incorrect photometry, and are tagged in
Table~\ref{src-list} with an `x'.  These sources have been omitted
from the final data set, leaving the total number of $JHK_s$ sources
at 337.  It is this data set of 337 sources that is used for the rest
of the analysis in this paper.  The color-color diagram for these
sources is shown in Figure \ref{fig-cc} and the color-magnitude
diagram in Figure \ref{cmd3}.

\subsection{The IRS2 O Star Binary} 
\label{IRS2}

\begin{figure}
\centering
\includegraphics[width=3.5in]{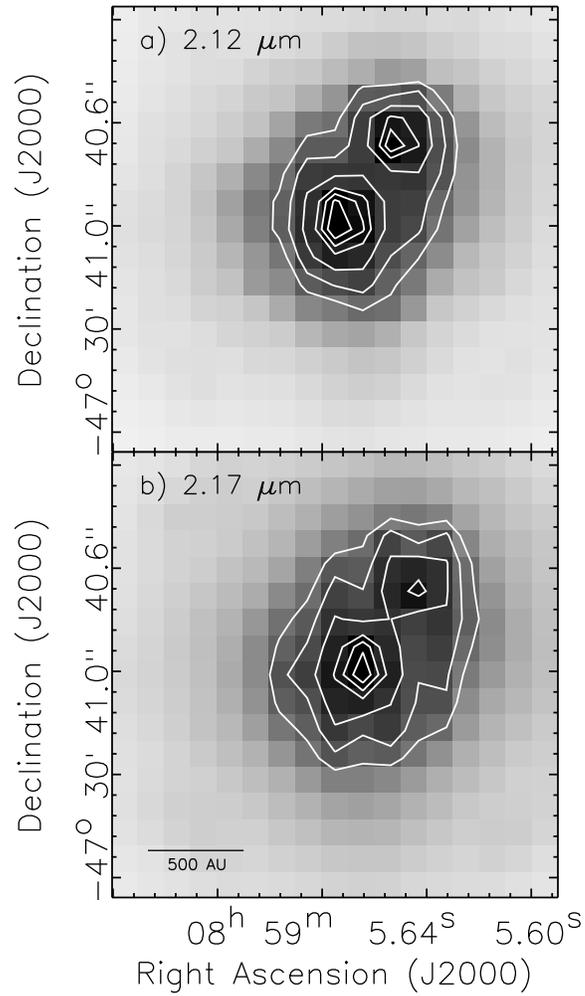}
\caption{a) Magnified image of IRS2 from the narrowband 2.12 \micron\
image.  Contours are plotted at 60, 70, 80, 90, 95 and 100\% of the
peak flux in arbitrary units. b) Magnified image of IRS2 from the
narrowband 2.17 \micron\ image.  Contour levels are the same as in a).
}
\label{irs2}
\end{figure}

IRS2, the central source of RCW~38, has been previously thought to be a
single source, but with the increase in resolution enabled by the AO
system, it is possible to show that it is in fact a binary system.  The
separation of the two sources can be confirmed in all three broad-bands
through visual inspection, but due to the brightness of the two sources,
the detector is saturated in this region for the broadband $JHK_s$ filters.
Therefore, the narrowband images (2.17\micron\ and 2.12\micron) have been
used for a relative photometric analysis of IRS2.  The ratio of the
detected flux from the brightest pixel in each source shows the dimmer
source to have $\sim94\%$ of the flux of the brighter source (89$\%$ in the
case of the 1.28\micron\ image).  Comparing this ratio to the ratio of
Lyman continuum photons given for various ZAMS spectral types in Panagia
(1973; Martins et al.\ 2005), shows that both stars in the binary should
have the same spectral type.  Figure \ref{irs2} shows magnified images of
IRS2 with contours in these two bands.  From the narrowband images, the
projected separation between the two sources is 0\farcs27 which is 459 AU
at 1.7 kpc.

Smith et al.\ (1999) address the question of the effective temperature and
hence spectral type of IRS2 by treating it as a single source and fitting
nebular spectroscopic lines to models of HII regions using stars with
Kurucz stellar atmosphere models, from which they determine an effective
temperature in the range of 43000K to 48000K.  Smith et al.\ favored the
lower range for IRS2, 43000 $-$ 44000 K, corresponding to a spectral type
of O5 $-$ O5.5 (Panagia 1973), but not ruling out a temperature of 48000 K,
corresponding to an O4.5 spectral type.  More recent modelling suggests the
lines used by Smith are very sensitive to model parameters, and after an
effective temperature of 40000K, the lines are rather insensitive (Mokiem
et al.  2004).  Therefore, a new look at Smith et al. suggests only that
the effective temperature is greater than 40000K.

The ionizing flux can be measured independently at radio wavelengths.
At 5 GHz, Wilson et al.\ (1970) measured a flux of 172 Jy (over a large area
with a 4\arcmin\ beam), while Vigil (2004) measured 99 Jy (with a
10\arcsec\ beam interferometrically) at the same frequency.  The Vigil
(2004) observations focused on the immediate region around IRS2, and is
likely to be more representative of the flux solely due to IRS2.  

Taking the electron temperature to be 7500 K (Caswell \& Haynes 1987), and
assuming the Wilson et al.\ and Vigil results represent the upper and lower
limits respectively, we find that the number of Lyman continuum photons,
$N_L$, is in the range $2.7 - 4.7\times10^{49}$ s$^{-1}$ (Armand et al.\
1996), or log ($N_L$) $\sim 49.1 - 49.4$.  Following Martins et al.\ (2005)
this range of log ($N_L$) values implies a spectral type of O4 to
$\sim$O5.5 for luminosity class (LC) V stars, and O5.5 to O7 for LC III
stars.  For an effective temperature $\geq$ 40000 K, the calibration of
Martins et al.\ (2005) suggests a spectral type of O5.5 or earlier for LC V
and O5 or earlier for LC III.  Therefore, the radio flux and effective
temperature measurements give inconsistent results for LC III, while for LC
V they are consistent.  Futher, as these are young stars, LC V is more
likely, and thus we conclude that the spectral types of the IRS2 binary,
which we assume to be of equal mass, are likely to be $\sim$O5 V, with a
range of O4-O5.5 consistent with the available data.  Note that the other
OB star candidates in the immediate vicinity of IRS 2 have log ($L_{bol}$)
$\leq$ 3.4 (Wolk et al.\ 2006), and thus are not likely to contribute
significantly to the continuum radiation.

\subsection{Cluster Membership}

Completeness histograms of $H$ and $K_s$ magnitudes do not show the
characteristic steep rise towards the fainter magnitudes that is often
present due to background sources if the cloud is being penetrated.  This
implies that the region is relatively free from extensive background
contamination to the depth of our survey (Carpenter et al.\ 1997), but the
likelihood of a small fraction of the detections being from background
sources is estimated from the SOFI control field.  

\begin{figure}
\centering
\includegraphics[width=3.5in]{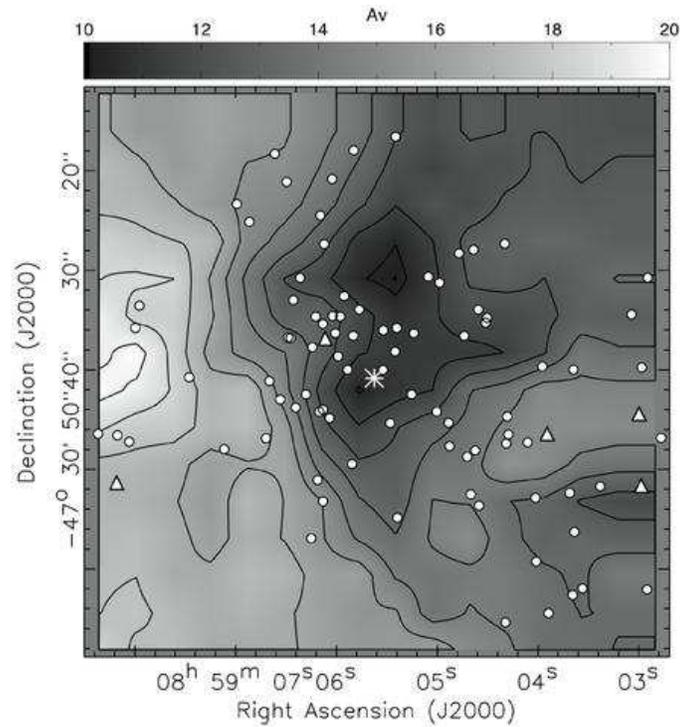}
\caption{Shaded extinction contour map for the NACO FOV. White areas
indicate areas of higher extinction, expressed in a linear scaling where
black represents $A_V=10$ and white represents $A_V=20$.  Contours are
given for $\Delta A_V=1$, starting at $A_V = 10$.  $K_s$-band excess
sources and potential protostellar candidates as identified in
Figure~\ref{fig-cc} are overplotted using the same symbols as that figure
(white circles and white triangles respectively).  The
bright central source IRS2 is overplotted as a white star.}
\label{extmap}
\end{figure}

An extinction map for the NACO field was calculated following the method of
Gutermuth et al.\ (2005) and is shown in Figure~\ref{extmap}. We used
$H-K_s$ colors for groups of the 20 closest sources detected in all bands,
sampled at each point in a uniform grid with a spacing of 3\farcs7, in
order to analyze the extinction surface density.  The reddening law of
Rieke and Lebofsky (1985) was used.  Sources used are all sources detected
in $H$ \& $K_s$ with $\sigma < 0.2$.  It is important to note that the
measurements used to make this extinction map are dominated by cluster
members, and so it is mostly a map of the extinction in front of and
intrinsic to the cluster.  

The NACO extinction map yields a mean extinction of $A_V=14.8$, which is a
much higher extinction than the SOFI field due to the molecular cloud.  The
average extinction in the SOFI $K_s$-band control field is $A_V=1.83$,
which was calculated as described above for the NACO field.  Therefore, for
comparison, the control field $K_s$-band magnitudes were artifically
extinguished by $A_K=1.45$ magnitudes.  This is an underestimate of the
reddening, thus the amount of contamination should be considered an upper
limit.  The extinguished sample yields a 90$\%$ completeness limit of 17.12
magnitudes for the control field compared to 16.97 for the RCW~38 field.

\begin{figure}
\centering
\includegraphics[width=3.5in]{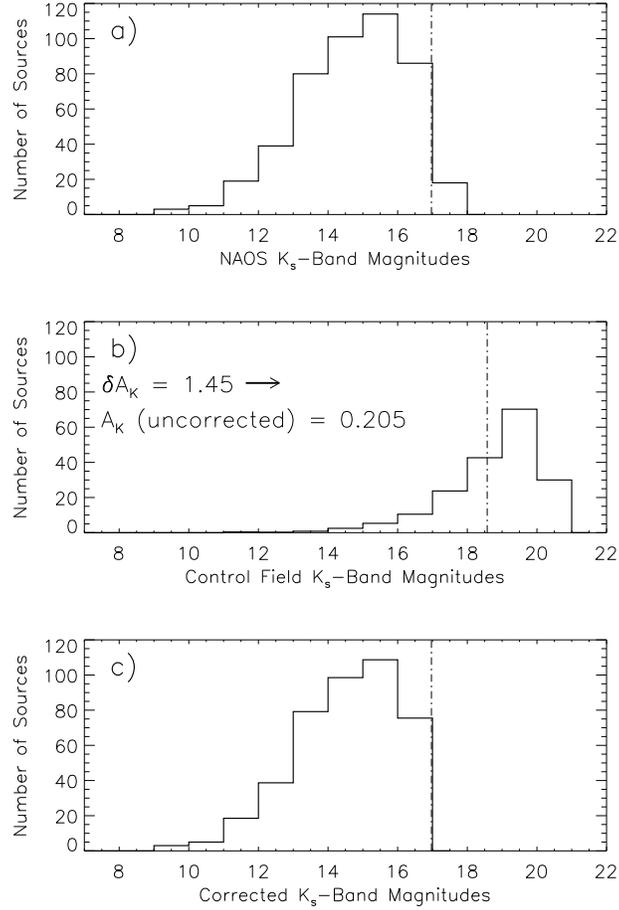}
\caption{a) Histogram of NACO $K_s$-band magnitudes divided into bins of 1
magnitude.  All sources detected at $K_s$, regardless of detectability at
$J$ or $H$, are included. The dotted line represents the $90\%$
completeness limit.  b) Histogram of SOFI control field $K_s$-band
magnitudes divided in bins of 1 magnitude and normalized by the ratio
between the NACO FOV and the SOFI FOV.  The dotted line represents the
$90\%$ completeness limit.  The histogram has been shifted to the right by
the amount $\delta$A$_K$ as described in the text.  c) NACO histogram
corrected by subtracting the SOFI control field histogram.  Only bins
within the $90\%$ differential completeness limit for the NACO measurements
are included.
}
\label{cfhist}
\end{figure}

Figure~\ref{cfhist} shows $K_s$ magnitude histograms (regardless of a
detection or lack of at $J$ and/or $H$) for the NACO RCW~38 data, the
artificially extinguished control field data scaled by area, and the
``corrected" histogram given by their subtraction.  The total FOV of the
control field is $4\farcm2 \times 4\farcm2$ and the area covered in the
NACO observations is only 5.8$\%$ of this, which gives the area scaling
factor.  Integrating over the bins brighter than the NACO 90$\%$
completeness limit cutoff and using $\sqrt{n}$ counting statistics to
calculate error, the histogram shows 424 total NACO $K_s$-band sources and
$N_{bkgd} = 20 \pm 2$ control field sources (scaled by the area ratio) so
that $4.7\%$ of the NACO sources are potential background contaminants.
Therefore, of the 337 $JHK_s$ sources (including protostellar candidates),
$317\pm5$ sources can be considered cluster members.  This estimate of
background contamination should represent an overestimate because the
number of reddened background stars is reduced by requiring $J$ and $H$
detections as well.  Because the contamination from background sources is
small, all 337 sources are included in the analysis and diagrams of this
paper, except for the disk fraction analysis where the corrected total of
$317\pm5$ sources is used.

The small amount of contamination is due to two reasons.  One is simply the
small field-of-view of our observations, combined with the extinction in
front of and intrinsic to the cluster, of $A_V \sim 15$.  The other is the
large background column density, as derived from observations of the
thermal dust continuum near 1-mm wavelength (Cheung et al.\ 1980).  The
peak column density is measured to be $8 \times 10^{23}$ cm$^{-2}$,
corresponding to an extinction of $A_V = 400$ (Wolk et al.\ 2006).  Thus it
appears that the embedded cluster lies in front of a dense molecular cloud
whose high extinction screens out background sources. 

\subsection{Candidate Protostars and $K_s$-band only Sources}
\label{sec-proto}

There are five $JHK_s$ sources whose 1$\sigma$ error bars in both
coordinates of color-color space fall completely outside the reddening
vector extending from the T-Tauri locus in Figure~\ref{fig-cc}.  These
are protostellar candidates identified in a similar way as Lada \&
Adams (1992), and they represent approximately 2$\%$ of the
population.  They are tagged in Table~\ref{src-list} with a `p'.
These candidates were visually inspected and appear to be true
detections.  Lada et al.\ (2000) find that protostellar candidates
compose 13$\%$ of sources detected in $JHKL$ in the Trapezium Cluster,
with the difference likely due to the additional $L$-band sensitivity
represented in the Lada et al. data set.

There are also 53 sources with magnitude error less than 0.2 that were
detected in $K_s$-band only.  These sources are mostly cluster members due
to the small amount of background contamination as shown in Figure
\ref{cfhist}.  These sources could either be highly extinguished T-Tauri
stars or protostellar candidates.  We have compared their distribution to
the extinction map shown in Fig~\ref{extmap}, and we find that most of the
sources are located in regions of high extinction.  In Figure~\ref{im}
these sources are clearly seen in the east and south-east as red sources.
There are also 65 sources with magnitude error less than 0.2 that were
detected in $H$ and $K_s$-band only.  These sources are distributed in a
similar way to the sources only detected in the $K_s$-band, suggesting they
are likely extinguished T-Tauri stars or embedded sources.

\subsection{Molecular Hydrogen Emission}

\begin{figure}
\centering
\includegraphics[width=6in]{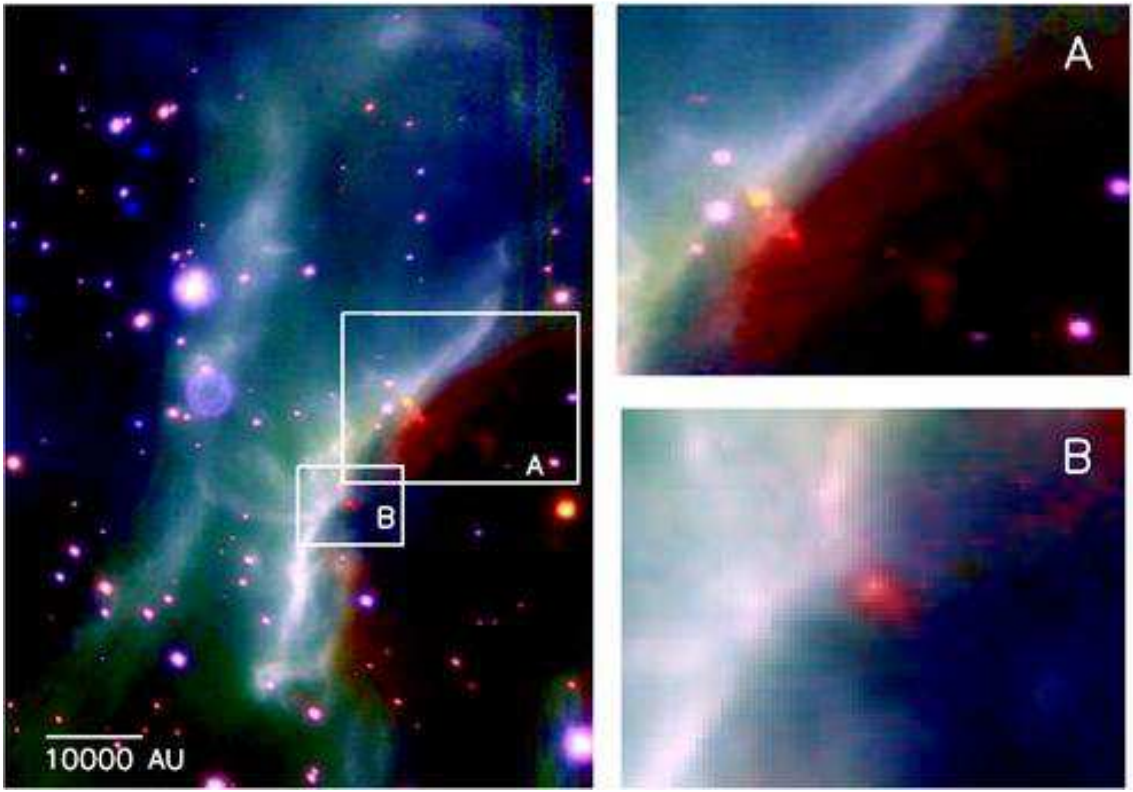}
\caption{Molecular hydrogen emission in the 1-0 S(1) line is revealed in
these images, which combine narrow-band images at 2.12 \micron\ (red),
2.17 \micron\ (green), and 1.28 \micron\ (blue).  In the left panel is
shown the IRS~1 ridge region, and on the right are shown zoomed in regions
A and B.  The H$_2$ emission is shown in red.}
\label{outflows}
\end{figure}

Narrowband imaging of the H$_2$ 1-0 S(1) line at 2.122 \micron\ is
often used to identify shocked molecular emission due to outflows.  In
a nebulous region such as RCW 38, pure H$_2$ line emission must be
distinguished from the extensive and bright continuum emission using
multi-band data.  By comparing the 2.122 \micron\ narrow-band image with the
broad-band $K_s$ image, two regions of H$_2$ emission have been
identified (Fig.~\ref{outflows}).  Both lie along the IRS~1 ridge,
which has previously been identified as a region of active star
formation (Smith et al.\ 1999; Wolk et al.\ 2006).  Region A in
Fig.~\ref{outflows} shows H$_2$ emission extending to the SW from an
infrared point source that is one of the redder sources in the region
(No.\ 174 in Wolk et al.\ 2006).  This infrared source is coincident
with one of the brighter X-ray sources (over 500 counts) but is barely
detected at $J$ in the NACO data.  It is likely to be an embedded
protostar with an outflow.

In region B, H$_2$ emission extends to the SW from a kink in the IRS~1
ridge emission, suggestive of a relationship between the two features.  No
infrared point source can be directly associated with this emission
however, and so its origin is unknown.  A deeply embedded protostar may be
waiting to be found in this region, perhaps through deep mid-infrared
imaging and/or through interferometric imaging at millimeter wavelengths.
Unlike the ONC (Kristensen et al.\ 2003; McCaughrean \& Mac Low 1997; Allen
\& Burton 1993), or a large number of small scale outflows from the cluster
members (Davis et al.\ 2008), there is no large scale H$_2$ emission in the
RCW~38 region that could be suggestive of a recent explosive event, or a
large number of small scale outflows from the cluster members.

\subsection{Photoevaporating Disks}

\begin{figure}
\centering
\includegraphics[width=6in]{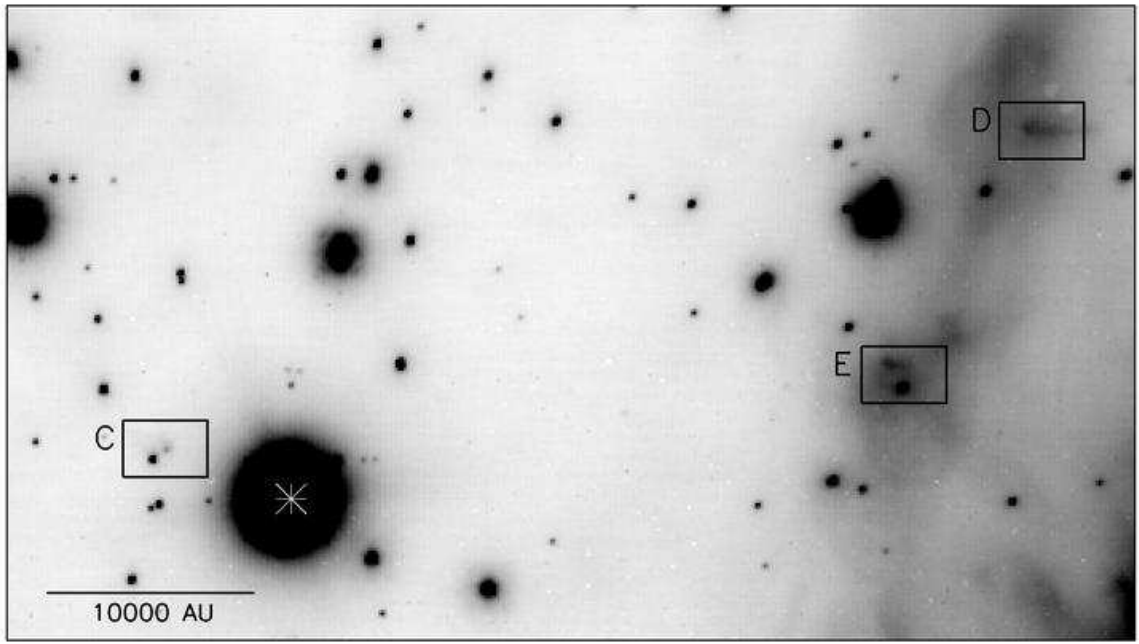}
\caption{
Finding chart for proplyd-like objects. The image is the narrowband 
2.17 \micron\ image, and the three proplyd-like objects shown in more detail
in Figs.\ \ref{proto_im1}-\ref{proto_im3} are indicated by the boxes marked
C, D and E.  IRS2 is indicated with an asterix.  
}
\label{proto_index}
\end{figure}

\begin{figure}
\epsscale{.9}
\plotone{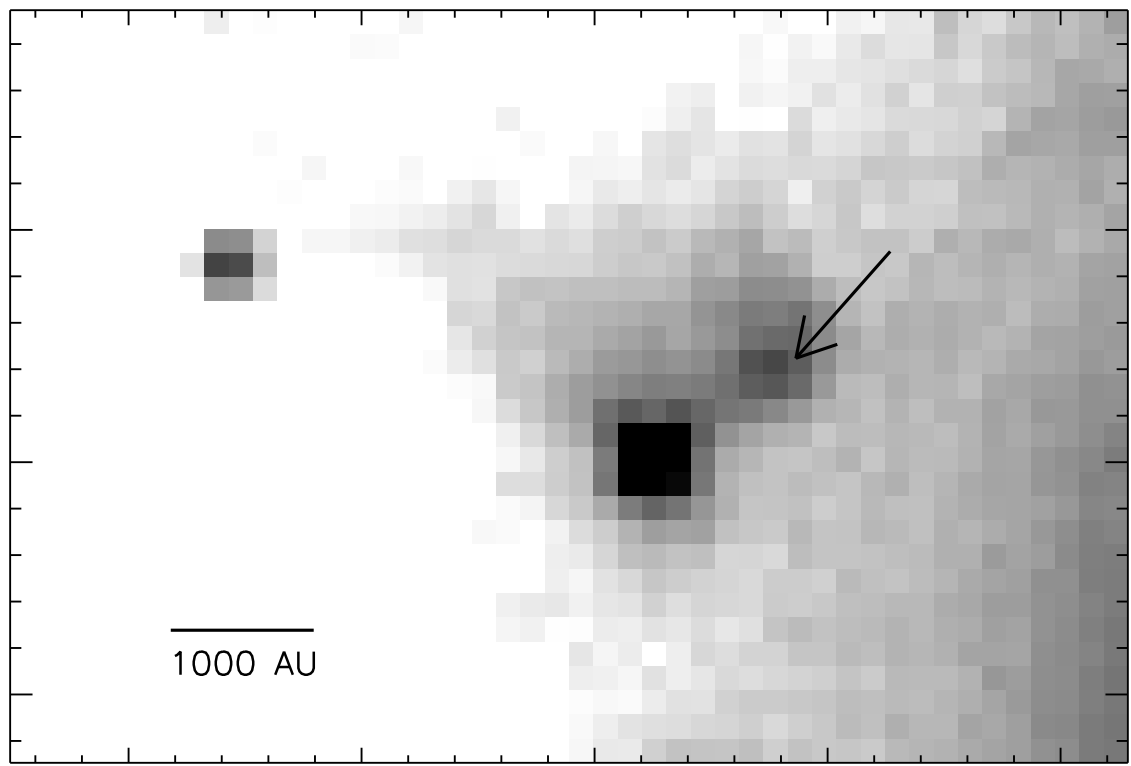}
\caption{Image at 2.12 \micron\ showing a candidate proplyd with a tail
near IRS2.  Identified as Region C in Figure \ref{proto_index}.  The arrow
points to the head of this globule, with the tail stretching leftwards and
slight upwards, away from the location of IRS2.
}
\label{proto_im1}
\end{figure}

\begin{figure}
\epsscale{.9}
\plotone{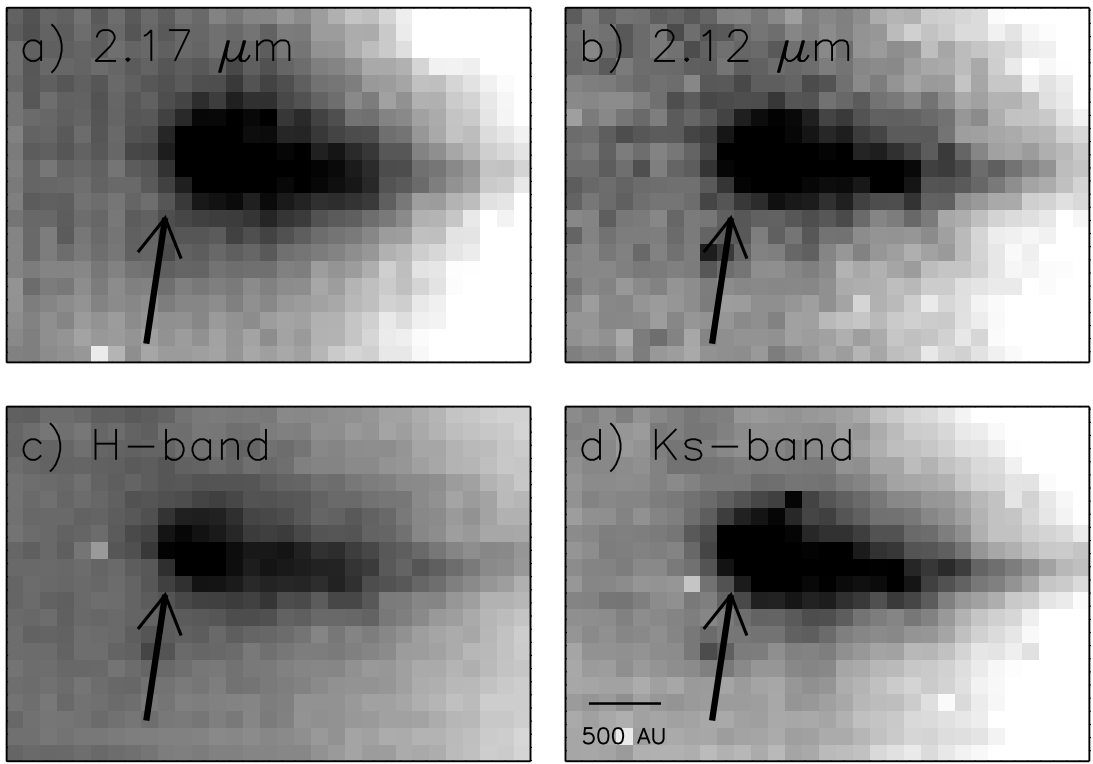}
\caption{Images in four bands (as noted) of a proplyd candidate with tail.
Identified as Region D in Figure \ref{proto_index}. 
}
\label{proto_im2}
\end{figure}

\begin{figure}
\epsscale{.9}
\plotone{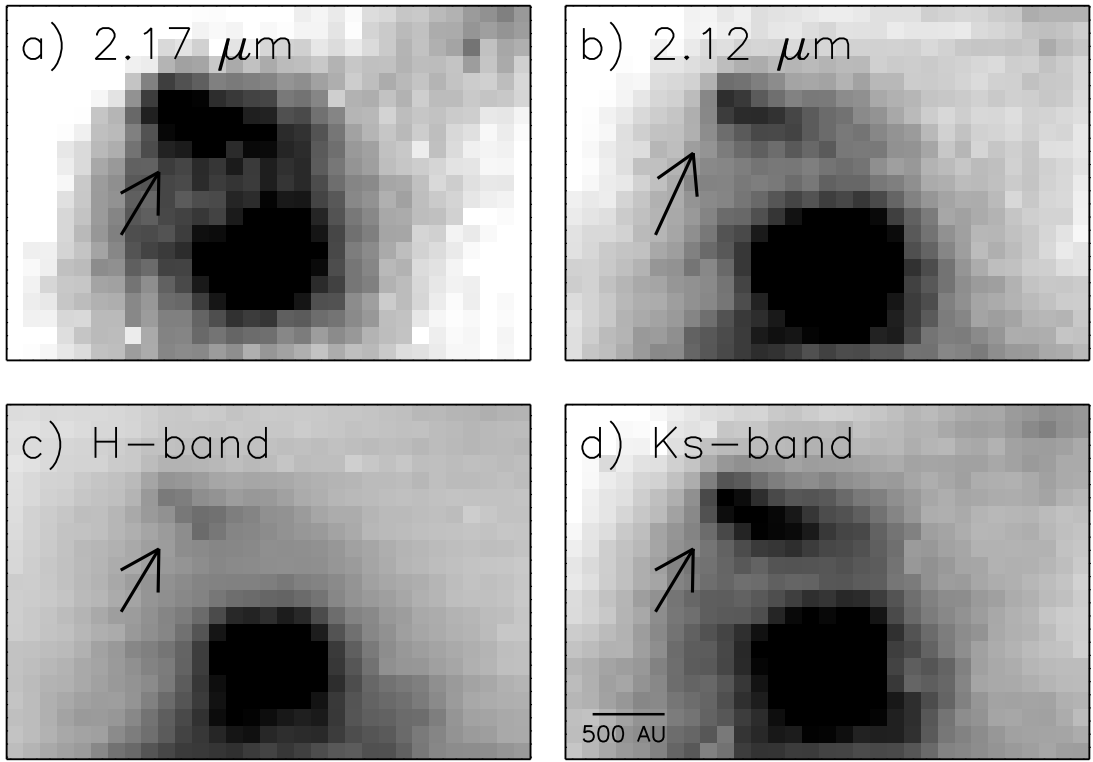}
\caption{Images in four bands (as noted) of a proplyd candidate with
tail.  Identified as Region E in Figure \ref{proto_index}.  The arrows
point to the head of the globule.}
\label{proto_im3}
\end{figure}

The narrowband imaging reveals three candidate photoevaporating disks --
stars with cometary-like head-tail morphologies (Figs.~\ref{proto_index},
\ref{proto_im1}, \ref{proto_im2}, \ref{proto_im3}).  While all have
stellar-like point sources at their heads, only one of these is clearly
directly affected by IRS 2, with its tail pointing directly away from the
O-star and lying in close proximity (Fig.~\ref{proto_im1}).  These objects
are well seen in the narrowband images at 2.12 \micron\ (H$_2$) and 2.17
\micron\ (Br$\gamma$), when compared to the broadband images at $H$ and
$K_s$ that are dominated by continuum emission.  The proplyds in Orion have
typical sizes 0\farcs15 to 1\arcsec\ (O'dell 1998), which translates to
62--413 AU assuming the distance to the ONC is 414$\pm$7 pc (Menten et al.\
2007).  Our observations have a pixel scale of 0\farcs054 (91 AU at a
distance of 1.7 kpc), which should allow for the identification of any
larger Orion-like proplyds in the region.  We have detected at most three
similar objects with tail sizes that are approximately 1275 AU (Fig.
\ref{proto_im2}) and 2208 AU (Figs. \ref{proto_im1}, \ref{proto_im3}).  It
is interesting to note that two of these objects are located in the IRS 1
ridge.  One of the goals of undertaking both narrowband and broadband
imaging was to identify both resolved and unresolved proplyds.  In the case
the proplyds are unresolved, they would be identified as sources of line
emission in the narrowband images after careful subtraction of the
continuum (broadband) emission.  However, the variable PSF, both in time
and across the FOV, make such a comparison difficult if not impossible, and
we have not attempted to do this.  Stable PSFs like those provided by
space-based platforms are needed to undertake a search for unresolved
emission from evaporating disks.

\subsection{Disk Fraction}

\begin{figure}
\centering
\includegraphics[width=3.5in]{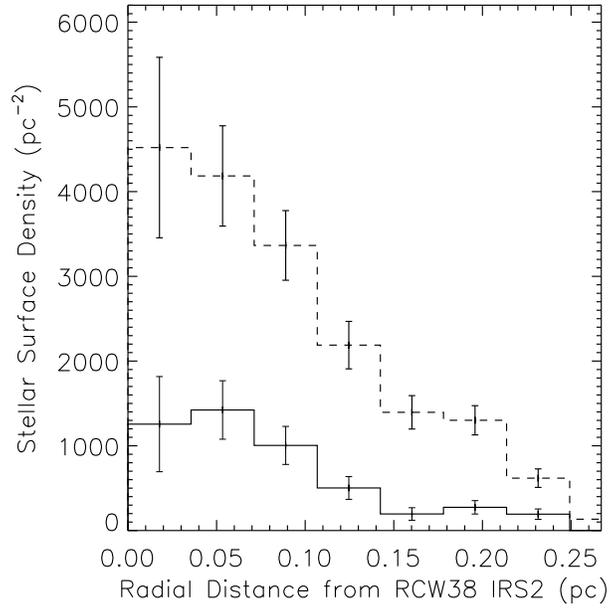}
\caption{Radial surface density plot IRS2.  Bin size is 4\arcsec.  
The dashed line indicates all $JHK_s$ sources and the solid line 
is $K_s$-band excess sources from Figure \ref{fig-cc}.
}
\label{raddist}
\end{figure}

\begin{figure}
\centering
\includegraphics[width=3.5in]{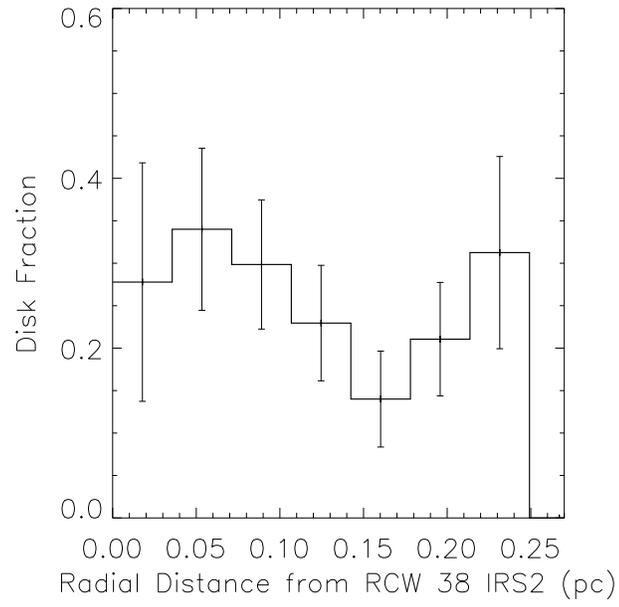}
\caption{Disk fraction vs. radial distance from IRS2. 
Bin size is about 4\arcsec.
}
\label{raddist_disk}
\end{figure}

Using the ($J-H$) vs. ($H-K_s$) color-color diagram (Figure
\ref{fig-cc}), it is possible to isolate 87 sources (92 including
protostellar candidates) with $K_s$-band emission greater than that
expected for a stellar photosphere.  These sources are tagged in
Table~\ref{src-list} with an `e' for excess sources, and a `p' for
protostellar candidate.  Figure \ref{extmap} shows a spatial
distribution of these sources and Figure \ref{raddist} shows a
histogram of sources vs. radial distance from IRS2.  Excess infrared
emission is often indicative of the presence of inner disks.  The disk
fraction $D$ is calculated as:

\begin{equation} 
D=\frac{N_{excess}}{N_{tot}-N_{bkgd}} 
\end{equation}

where $N_{excess}$ is the number of sources with an infrared excess,
$N_{tot}$ is the total number of sources, and $N_{bkgd}$ is the number of
background sources.  The disk fraction as a function of radial distance
from IRS2 is plotted in Figure~\ref{raddist_disk}.  In this manner we find
a disk fraction of $29 \pm 3\%$.  The disk fraction is quoted without mass
and extinction limited sampling because it did not affect the final value.
We perform mass and extinction limited sampling when comparing RCW~38 and
the ONC.  This is a lower fraction than found in similar studies ($JHK$) of
similar clusters (Lada et al.\ 2000; Haisch et al.\ 2000; Hillenbrand
2005), given its degree of embeddedness ($A_V \sim10$) and inferred young
age (0.5 Myr; Wolk et al.\ 2006).

As mentioned previously, RCW~38 is an interesting region to study in that
we are able to compare what we find to the best studied young cluster, the
ONC, to see if the trends found there exist elsewhere.  Lada et al.\ (2000)
in their $JHKL$ survey of the ONC covering approximately the same linear
scale as our RCW~38 study (a 0.4 pc square region) found a disk fraction of
$50\pm7\%$ determined by $K_s$-band excess, while a disk fraction of
$80\pm7\%$ was determined using $L$-band excess.  Both $K_s$ and $L$ are
sensitive to warm inner disk emission ($\leq$ 1 AU) and require favorable
angles of inclination, inner disk holes, and accretion rates for detection
(Lada et al. 2000).  However, the photospheric emission of a given star has
a lower intensity in $L$ rather than $K_s$, which means the $K$-band excess
must compete with photospheric emission much more than $L$-band.  Thus,
longer wavelengths such as the L-band, that can more reliably detect disks,
resulting in higher disk fractions than for studies relying on $K_s$-band
excess only, are needed.  It is highly likely that our survey of RCW~38
only detects a fraction of disks present.  The only way to test this would
be to image the region in $L$-band or longer wavelengths with sufficient
sensitivity and resolution to resolve individual sources, which is a
difficult task for a cluster at 4 times the distance of the ONC (1.7 kpc)
with bright nebulosity.

The disk fraction of $50\pm7\%$ found for the ONC using $K_s$-band excess
is difficult to directly compare to the present result due to differences
in sensitivities and photometric techniques.  In order to make a fairer
comparison, the $JHK_s$ data on the ONC from Muench et al.\ (2002) was
obtained and a 0.3 M$_\sun$ mass cut and a ($J-H$) $<$ 2 extinction cut
applied to both data sets, over the same linear region.  This resulted in
218 sources in the ONC, including 71 excess sources, and 112 sources in
RCW~38, with 28 excess sources.  The overall disk fractions for the two
data sets subjected to these cuts are $25 \pm 5\%$ for RCW 38 and $42 \pm
8\%$ for the ONC. These fractions are significantly different, for reasons
unknown. As noted, however, determining the infrared excess fraction from
$JHK$ data only is likely to underestimate the true fraction (Lada et al.\
2000).  

\section{Analysis}

\subsection{Photoevaporating Disks?}

Sources with disks in RCW~38 seem to be evenly distributed throughout the
region, with the majority of disks found between 0.07 and 0.17 parsecs from
IRS2 (48\%, compared with only 24\% at smaller distances).  Because
of the UV radiation emanating from the massive O-star binary IRS2, it might
be expected that the disks nearest to the central source would have been
evaporated.  Photoevaporating disks in the inner region of a cluster have
been well documented in the Trapezium Cluster on account of the O star
$\theta^1$ C Ori through both radio observations (Churchwell et al.\ 1987;
Felli et al.\ 1993) and both ground and space based optical imaging (Laques
$\&$ Vidal 1979; O'dell, Wen, $\&$ Hu 1993).  Instead of a deficit of disks
near IRS2, we find a slight increase in disk fraction with decreasing
projected distance from IRS2 (Fig~\ref{raddist_disk}).  However, the
physical separations are not known and the proximity of some of these stars
to IRS2 could be a projection effect.  In addition, since the cluster is
quite young and thus dynamically active, stars moving at 1 km/s would move
across the entire FOV in just 0.5 Myr, making it nearly impossible to tell
which stars have spent prolonged amounts of time near IRS2.  This could
explain the apparent lack of resolved proplyd-like objects in the inner
region near to IRS2.

It is also possible that the outer disks of many of these sources have been
blown away, leaving the inner disks intact which is the portion most
sensitive to $K_s$-band measurements.  O5.5 stars such as IRS2 emit large
quantities of Far-Ultraviolet (FUV) and Extreme-Ultraviolet (EUV).  More FUV
radiation is emitted than EUV radiation, which means that FUV radiation is
the dominant cause of photoevaporation because it works to evaporate the
outer portions of the disks at a faster rate.  EUV radiation works to
evaporate the disk at a slower rate, so when the outer portion of the disk
has been photoevaporated from FUV radiation, it is the more energetic EUV
radiation that slowly works to cook away the inner disk.  Adams et al.\
(2006) modeled this photoevaporation effect for embedded clusters with
varying memberships.  They define a $G_0$ term given in units of the
typical interstellar radiation field at FUV wavelengths (given as $1.6
\times10^{-3}$ ergs s$^{-1}$ cm$^{-2}$).  For an O star binary (assuming a
temperature of 41000K and $R_\star=15 R_\sun$), sources at 0.01 pc would
experience $G_0 = 43000$ which decreases as 1/(distance)$^2$ so that
sources at 0.1 pc would experience $G_0 = 430$.  Since the O stars
represent the dominant source of FUV radiation, the $G_0$ values given take
into account just the central source and ignore contributions from nearby
less massive sources.  However, if we were to take into account the 31
candidate OB stars identified in Wolk et al.\ (2006), the average FUV
radiation experienced by a given star would be greater.  Higher $G_0$
values represent a faster photoevaporating effect and most easily affect
lower mass stars whose disks are less tightly bound.

If we assume large planet formation occurs in the region of 5-30 AU, for
$M_\star=0.5 M_\sun$ or less it would be inhibited over a 2 Myr period in a
region where $G_0=3000$, though this would only impact the stars minimally
if the cluster is $<$ 2 Myr.  As an example from Adams et al.\ (2004),
disks around 0.25 $M_\sun$ stars would be photoevaporated down to around 14
AU, reducing the reservoir of material for planet formation around the star
by half and likely preventing the formation of outer planets.  Without
velocity dispersion data for the cluster, however, it is difficult to know
what kind of orbits the cluster members follow around IRS2 and how much
time they actually spend close to this central source.  However, if we
assume circular orbits for cluster members so that they spend $100\%$ of
their time at their projected radial distance, there are 18 $JHK_s$
sources that experience $G_0\geq3000$.  Of these, 5 sources have detected
infrared excess, giving a disk fraction of 27$\%$.  However, random
sampling of the sources using the disk fraction calculated previously gives
a 28$\%$ chance of choosing 5 or less excess sources, so that this slightly
lower disk fraction is not statistically significant.  This result is to be
expected since $K_s$-band excess detects the inner wall of disks which can
be found at distances less than 1 AU from the central source so that even
if a star's disk was significantly reduced in size, this lack of disk
material would not be noticeable.  As an example, Eisner et al.\ (2008)
measured the mass of disks in the ONC versus Taurus and found that the ONC
disks are much less massive, indicating that photoevaporation effects from
the OB stars in the field, have played a key role in evaporating the outer
disks in the ONC.

\subsection{Cluster Structure}

\subsubsection{Azimuthally Averaged Surface Density Profile} 

The azimuthally averaged surface density profile, or radial profile, of a
cluster is a common tool used to probe the decrease in surface density as a
function of radius from the central peak (Carpenter et al.  1997;
Hillenbrand \& Hartmann 1998).  King (1962) surveyed 15 globular clusters
in order to examine their radial profiles and found a relationship that
could be expressed with only the tidal radius and the core radius of a
cluster.  Hillenbrand \& Hartmann (1998) fitted their radial profile for
the ONC using such a King profile.  Neither the ONC or RCW 38 is a globular
cluster in equilibrium which makes the King model less physically
significant for this application; however, as RCW 38 shares many
similarities with the ONC, it seemed reasonable to determine the King
profile parameters for RCW 38 for comparison to the Hillenbrand \& Hartmann
values. 

\begin{figure}
\centering
\includegraphics[width=3.5in]{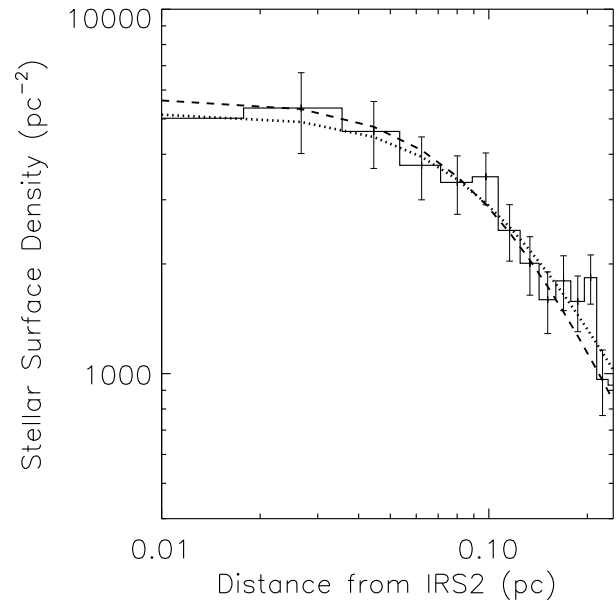}
\caption{Radial stellar surface density plot for RCW~38 centered on IRS2
using NACO data.  Bin size is about 0.02 pc (2\farcs5).  The stellar
surface density is computed for all sources detected with $K < 17$ with
$\sigma<0.2$.  The dashed line represents the fitted King model
(eqn.~\ref{eqn-king}) while the dotted line represents the fitted Elson et
al.\ (1987; EFF) model (eqn.~\ref{eqn-eff}).
}
\label{radplot1}
\end{figure}

\begin{figure}
\centering
\includegraphics[width=3.5in]{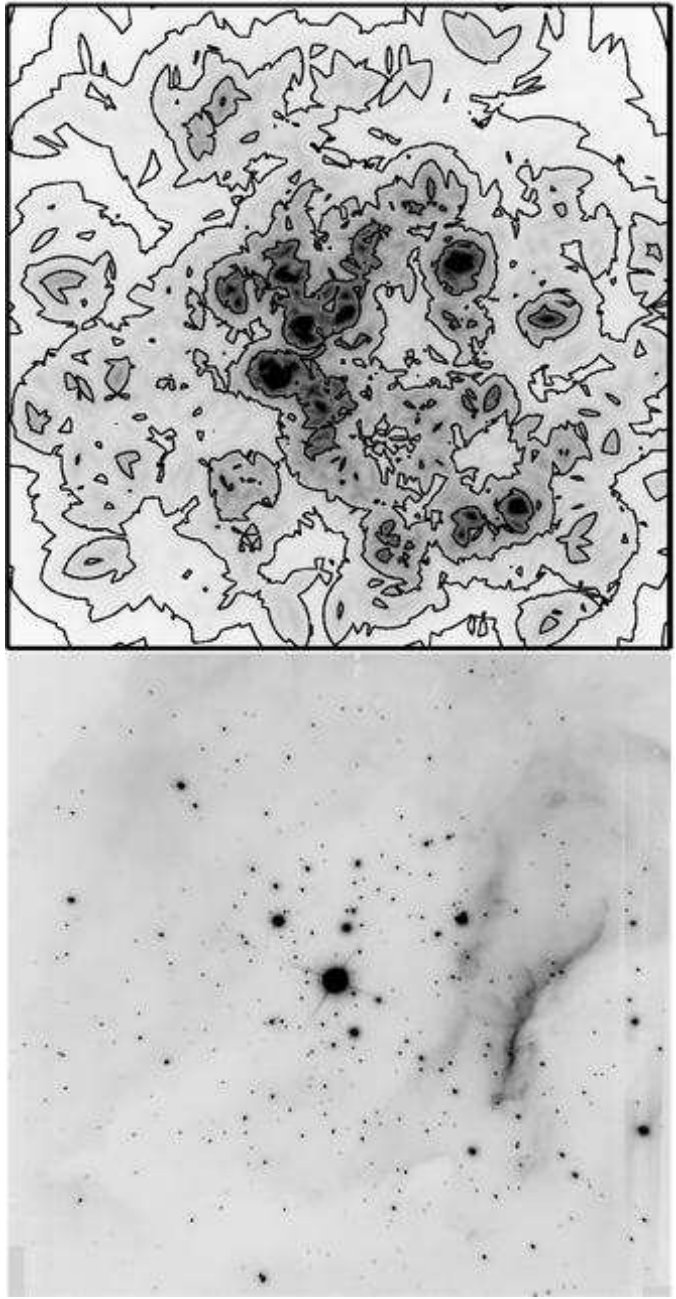}
\caption{Stellar surface density plot.  The contours of $K_s$ stellar
surface densities for sources with $K < 17$ and $\sigma<0.2$ are shown in
the upper image.  The contours represent stellar surface densities of 30,
100, 500, 1000, 2500, 5000, 10000, 20000, and 25000 stars pc$^{-2}$.
Densities are computed using 5 nearest neighbors.  For reference, in the
lower panel is shown the same FOV using the $K_s$ band image.
}
\label{contourjhk}
\end{figure}

The radial profile of RCW~38 was made using only the sources detected
at $K_s$ with magnitude $<$ 17 and uncertainties $<$ 0.2. As shown in
Figure \ref{radplot1} it is relatively flat inside the core of the cluster
at radial distances less than 0.1 pc and then begins to fall off at further
distances.  The radial profile can only be examined up to a distance
of 0.25 pc from the center of the cluster in order to limit edge effects
caused by sampling distances outside the FOV in certain radial directions.
This radial profile represents only the sources resolved in this sample,
whereas it is possible that there are deeply embedded and extinguished
sources that are cluster members but not taken into account in this radial
profile.  Figure \ref{contourjhk} shows a contour map of $K_s$ stellar
surface density that illustrates the fairly uniform distribution with
only a few small areas of possible subclustering.

The shape of the radial profile in Figure \ref{radplot1} motivates the
choice of a flat-topped profile with a power-law drop off to fit the
radial profile of the data such as the King model.  For regions where
the tidal radius is much larger than the core radius (as is generally
assumed), the King model can be expressed as:

\begin{equation} 
\sigma(r)=\sigma_0\left[1+\left(\frac{r}{r_0}\right)^2\right]^{-1}
\label{eqn-king}
\end{equation}

where $r_0$ is the core radius of the cluster, and $\sigma_0$ is the
initial stellar surface density.  With $r_c=0.1$ pc and $\sigma_0=5700$
pc$^{-2}$, the model provides a reasonable fit to the radial profile of
RCW~38 (Fig.~\ref{radplot1}).  Hillenbrand \& Hartmann (1998) give best-fit
King model parameters of $r_{o}=0.164$ pc and $\sigma_{o}=5600$ pc$^{-2}$
for the ONC.  This central density is very close to that of RCW~38 while
the scale radius is larger, suggesting that RCW~38 is more centrally
concentrated than the ONC.

Because the King model is applicable for globular clusters, observations
show that it is not a good fit to the extended outer envelopes of young
star clusters in our Galaxy and the LMC and SMC.  In this case a modified
King model (Elson, Fall, \& Freeman 1987; EFF) is a better description:

\begin{equation}
\sigma(r)=\sigma_{o}\left[1+\left(\frac{r}{r_{o}}\right)^2\right]^{\frac{-\gamma}{2}}
\label{eqn-eff}
\end{equation}

where the introduction of the exponent $\gamma$ allows the more extended
nature of Galactic clusters to be fitted.  This model differs from the King
model in that there is no truncation radius included.  For RCW~38 we obtain
a best fit with $r_{o}=0.1$ pc, $\sigma_{o}=5200$ pc$^{-2}$ and
$\gamma=1.7$.  As in the case of the King model, the EFF model
provides a reasonable fit to the data in the flat and power-law fall-off
regions.

\subsubsection{Nearest Neighbor Distance Analysis}

\begin{figure}
\centering
\includegraphics[width=3.5in]{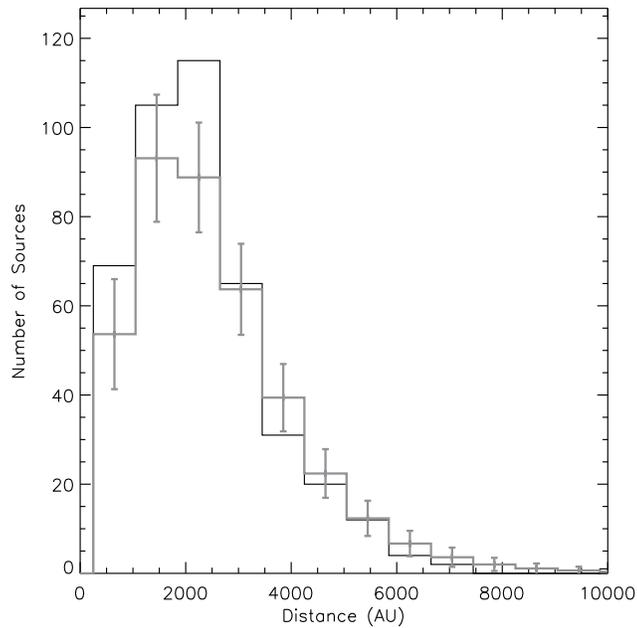}
\caption{Nearest neighbor histogram.  Black represents NACO data (using
sources with $K < 17$ and $\sigma<0.2$) and grey
is the average of 1000 randomly generated clusters with the same
radial profile as shown in Figure \ref{radplot1}. Error bars represent the
standard deviation of the random averaged clusters in each bin.  Bin size
is 800 AU.  As a note, the two-point correlation function (TPCF) of this
same data set yields nearly identical results, so it is not included here.
Since the data does not deviate significantly from the model (where sources
are randomly distributed for the given radial profile), it is consistent
with the model that no major substructuring exists in this FOV.  This is as
expected because the radial density plot in Figure \ref{radplot1} is
essentially flat over most of the region.
}
\label{nnhist2}
\end{figure}

We use a nearest neighbor distance analysis to look for subclustering
among close pairs.  Our King model (eqn. \ref{eqn-king}; dashed line in
Figure \ref{radplot1}) was used to generate a group of 1000 random
clusters with the same radial distribution and number of sources as RCW~38.
If the RCW~38 distribution of close pairs was significantly different from
the Poisson noise models, then we would expect subclustering among close
pairs.  A nearest neighbor distance histogram of the NACO data with a bin
size of 800 AU is compared with the averaged histogram values for these
model clusters in Figure \ref{nnhist2}.  The model fits with the data quite
well, with almost every bin matching when taking into account the error
bars.  This shows that there is a fairly low degree of subclustering on a
smaller scale, a fact that should be expected given the fairly flat radial
profile.

\subsubsection{Q Parameter Analysis}

Cartwright \& Whitworth (2004) use a $Q$ parameter in order to
classify star clusters based on the degree a cluster is dominated by
either fractal subclustering or a monolithic radial density gradient.
They show that a uniform 2D (or unconstrained 3D projected into 2D)
distribution gives $Q\sim$0.72 while a uniform 3D distribution
constrained to a spherical volume and projected into 2D gives
$Q\sim$0.79.  Thus, spherical clusters with some degree of
subclustering are expected to have $Q <$ 0.79 while clusters with a
smooth radial gradient are expected to have $Q >$ 0.79.

The parameter $Q$ is defined as $Q =\frac{m}{s}$ where $m$ is the mean
distance from a minimum spanning tree scaled by a factor of
$\frac{N-1}{\sqrt{NA}}$ (with $N$ being the number of stars and $A$
being the 2D cluster area) and $s$ is the mean of the two point
correlation function measured for all of the stars in the cluster
divided by the cluster radius. Using this procedure in order to
further characterize the structure of RCW~38, a value of $Q=$0.84 is
found, which places it on the border between clusters that are
characterized by a relatively uniform distribution (smooth with a
large scale radial gradient) and clusters that show some degree of
subclustering.  Together with the analysis in the previous section,
this result suggests that RCW 38 is quite uniform and does not show
strong subclustering.

\subsubsection{Larson Analysis}

\begin{figure}
\centering
\includegraphics[width=3.5in]{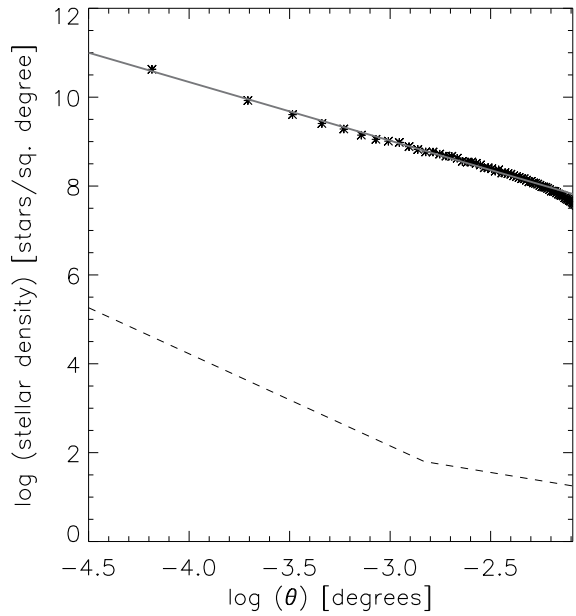}
\caption{
Surface density of companions as a function of angular separation (after
Larson 1995).   The Larson result, scaled to the same distance, is shown as
a dashed line (the original Larson result shows an elbow at around log
$\theta$ = -1.75 which translates to log $\theta$ = -2.83, for the distance
to RCW 38 of 1.7 kpc, cf. 140 pc for Taurus). 
The grey line shows a power law of -1.32 which is significantly different
from the -2.15 power law for the binary regime and -0.62 power law for the
clustering regime measured in Larson (1995). 
}
\label{larson}
\end{figure}

Figure \ref{larson} recreates the surface density vs. angular separation
plot from Larson (1995).  The power law for RCW~38 differs fairly
significantly from the original Larson plot and does not contain the same
elbow evident in the Larson plot that is identified as the transition from
the clustering to the binary regime.  The absence of the elbow is easily
explained by the fact that it occurs at an angular separation of 8663 AU,
which is at the very high end of the nearest-neighbor histogram in Figure
\ref{nnhist2} and thus not easily measurable by this smaller field of view.
Bate et al. (1998) also explains that we should expect the elbow in the
Larson plot to be washed out in a large 3D cluster where the angular
separation of many sources can be attributed to projection effects.  In
addition, from Figure \ref{larson} it is clear that RCW~38 is much denser
than the Taurus cluster and thus it is possible that different physical
processes govern star formation in these regions.

Nakajima et al. (1998) uses a similar Larson analysis on five young star
forming regions observed at visible wavelengths, fitting broken power laws
to their sample.  As a similar break in the Larson plot is not observed in
our RCW 38 data, due to the high density, we can only compare the slope
found at larger angular separations.  Table \ref{larsontab} shows the
slopes from Nakajima et al. (1998), where $\gamma_S$ gives the power-law
index at small angular separations while $\gamma_L$ gives the power-law
index at large angular separations, alongside the Larson result for Taurus
and the value for RCW~38.  The RCW~38 result is intermediate between the
values of $\gamma_S$ and $\gamma_L$ listed for other regions.

It is difficult to compare the deep near-infrared data from RCW~38 to the
$V$ and $R$ band data presented by Nakajima et al. because in those bands
there is poor completeness due to the high levels of extinction in the
visible band ($A_V\sim$ 10-30).  Because of the variability in extinction
and distance among the clusters surveyed, the mass sensitivities in each
region are likely to differ significantly.  Allen et al. (2002) show that
these factors can affect the Larson plot results significantly in their
HST/NICMOS survey of the Ophiuchus cluster.

\section{Summary}

We have presented high resolution adaptive optics near-infrared
imaging (broadband $J$, $H$, $K_s$; narrowband 1.28 \micron, 2.12
\micron, and 2.17 \micron) of the deeply embedded massive cluster
RCW~38, over a field-of-view of $\sim$ 0.5 pc, centered on the O5.5
star IRS2.  Our main results are:

\begin{enumerate}

\item A total of 360 sources are detected in all three broadbands.  Of
these, 344 have uncertainties in all bands of $<$ 0.2 mag, and the
remainder are removed from the sample.  A further seven sources were
removed due to contamination from other brighter nearby stars or due to
image artifacts, leaving 337 reliable sources.  Of these 337, five have
$H-K_s$ colors in excess of that expected for T Tauri stars, and are
considered to be protostellar candidates.  A comparison to a nearby control
field suggests that 15-25 sources are unassociated with the cluster.  There
are 53 $K_s$ only detected sources, that are most likely to be cluster
members due to the small amount of background contamination.  These sources
could either be highly extinguished T-Tauri stars or protostellar
candidates.

\item A $J-H$ vs. $H-K_s$ color-color diagram is used to identified stars
with an infrared excess indicative of inner disk emission.  In this manner
the disk fraction is found to be $29 \pm 3\%$.  For comparison to similar
data from the ONC, a 0.3 M$_\sun$ mass cut and a ($J-H$) $<$ 2 extinction
cut was applied to both regions over the same linear area, resulting in
disk fractions of $25 \pm 5\%$ for RCW 38 and $42 \pm 8\%$ for the ONC.

\item The central star IRS2 is found to be an equal mass binary of spectral
type $\sim$O5 and projected separation 459 AU.  

\item Molecular hydrogen line emission, a good tracer of shocked
outflow emission, is seen at two locations, from a comparison of the
narrowband 2.12 \micron\ images with both broadband $K_s$ and
narrowband 2.17 \micron\ images.  At one location this emission has a
morphology suggestive of outflow emission, and is associated with a
deeply embedded source that is also a hard X-ray source (Wolk et al.\
2006).

\item Three candidate photoevaporating disks are identified, with head-tail
morphologies, and with stellar-like sources at their heads.  Only one lies
in close proximity to IRS2 and has a tail pointing directly away from the O
star.  Although one goal of the study was to identify unresolved proplyds
through a careful comparison of narrowband and broadband imaging, this was
not possible due to the varying PSF between the bands.  

\item The cluster radial density profile is well described by a modified
King profile, which is widely used to fit globular clusters.  In young star
clusters, such a profile breaks down in fitting the extended outer
envelopes.  This is not a problem for the data presented here, as the FOV
is too small to trace the outer envelope region ($\sim$ 0.25 pc).  Nearest
neighbor analysis, both directly on the data and on a model density profile
that fits the data, show essentially no substructure (sub-clustering).
Comparison to other clusters suggests that this is not surprising for our
limit FOV.  The stellar surface density in the inner region of the
clusters is $\sim$ 5700 pc$^{-2}$.

\end{enumerate}

There are a number of observational programs on RCW 38 that need to be
undertaken in order to improve on this work and that of Wolk et al.\
(2006).  The comparatively low disk fraction of $\sim29\%$ found using only
$JHK_s$ data needs to be confirmed through sensitive observations in the
$L$-band.  The large distance and nebula emission will make these
observations difficult.  The nature of the protostellar candidates and
$K_s$-band only sources should be further investigated, through sensitive
mid-infrared imaging.  The OB star candidates identified in Wolk et al.\
(2006) should be observed with infrared spectroscopy to determine their
spectral types.  Finally, the age of the cluster needs to be better
determined, a difficult task given its extinction and youth.

\acknowledgments

We thank Joana Ascenso for providing the SOFI data in advance of
publication, and for sharing her ISAAC results in advance of publication.
We thank the VLT Science Operations team for successfully executing our
NACO observations in service mode.  Partial support for this work was
provided by NASA through contract 1279160 issued by JPL/Caltech, and by
NASA contract NAS8-03060.  This research has made use of NASA's
Astrophysics Data System.

Facilities: \facility{VLT:Yepun}

%% TABLES

\clearpage
\begin{deluxetable}{c|cc}
\tablecolumns{2} 
\tablewidth{0pt} 
\tablecaption{DITs (Detector
Integration Times) and NDITs for observations in each band.}
\tablehead{\colhead{Band} & \colhead{DIT (s)} & \colhead{NDIT} } 
\startdata 
$J$ & 4.0 & 16 \\ 
$H$ & 0.75 & 80 \\ 
$K_s$ & 0.50 & 120 \\
NB\_2.12 & 10.0 & 14 \\
NB\_2.17 & 10.0 & 12 \\
NB\_1.28 & 90.0 & 4 \\
\enddata
\label{dit}
\end{deluxetable}

\clearpage
\begin{deluxetable}{c|cc}
\tablecolumns{2}
\tablewidth{0pt}
\tablecaption{Number of detected sources.}
\tablehead{\colhead{Band} & 
\colhead{Detections with 0.2 magnitude uncertainty cap} }
\startdata
$J$ & 355 \\
$H$ & 420 \\
$K_s$ & 465 \\
$JHK_s$ All & 344 \\
\enddata
\label{sources}
\end{deluxetable}

\clearpage
\begin{deluxetable}{lllccccccl}
%\tablewidth{0pt}
\tablewidth{519.03049pt}
\tablecaption{Infrared Photometry of sources detected with magnitude
uncertainties $<$ 0.2.}
\tablehead{
\colhead{Source Number} & \colhead{R.A. (J2000)} & \colhead{Dec. (J2000)} &
\colhead{$J$} & \colhead {$J_{\rm err}$}  &
\colhead{$H$} & \colhead {$H_{\rm err}$}  &
\colhead{$K_s$} & \colhead {$(K_s)_{\rm err}$}  & 
\colhead{Notes} \\
}
\startdata
1 & 8 59 2.49 & -47 30 20.7 & \nodata & \nodata & \nodata & \nodata & 14.09 & 0.06 &  \\
2 & 8 59 2.50 & -47 30 52.6 & \nodata & \nodata & \nodata & \nodata & 15.02 & 0.09 &  \\
3 & 8 59 2.55 & -47 30 39.9 & \nodata & \nodata & \nodata & \nodata & 14.41 & 0.06 &  \\
4 & 8 59 2.56 & -47 30 46.2 & \nodata & \nodata & \nodata & \nodata & 12.08 & 0.02 &  \\
5 & 8 59 2.57 & -47 30 53.5 & \nodata & \nodata & \nodata & \nodata & 13.09 & 0.03 &  \\
6 & 8 59 2.60 & -47 30 31.7 & \nodata & \nodata & \nodata & \nodata & 13.35 & 0.04 &  \\
7 & 8 59 2.62 & -47 30 35.2 & \nodata & \nodata & \nodata & \nodata & 12.86 & 0.03 &  \\
8 & 8 59 2.63 & -47 30 28.7 & \nodata & \nodata & \nodata & \nodata & 15.07 & 0.08 &  \\
9 & 8 59 2.66 & -47 30 42.3 & \nodata & \nodata & \nodata & \nodata & 13.52 & 0.04 &  \\
10 & 8 59 2.67 & -47 30 57.4 & \nodata & \nodata & \nodata & \nodata & 15.68 & 0.14 &  \\
11 & 8 59 2.67 & -47 30 21.1 & \nodata & \nodata & \nodata & \nodata & 14.83 & 0.08 &  \\
12 & 8 59 2.74 & -47 30 43.4 & 17.72 & 0.07 & \nodata & \nodata & 15.32 & 0.09 &  \\
13 & 8 59 2.76 & -47 30 53.8 & 17.28 & 0.13 & \nodata & \nodata & 14.95 & 0.07 &  \\
14 & 8 59 2.78 & -47 30 46.8 & 14.97 & 0.02 & 13.61 & 0.02 & 12.65 & 0.03 & e \\
15 & 8 59 2.83 & -47 30 34.0 & \nodata & \nodata & \nodata & \nodata & 16.49 & 0.15 &  \\
16 & 8 59 2.89 & -47 31 6.5 & \nodata & \nodata & \nodata & \nodata & 15.25 & 0.10 &  \\
17 & 8 59 2.91 & -47 30 30.8 & 16.97 & 0.05 & 15.11 & 0.05 & 13.83 & 0.04 & e \\
18 & 8 59 2.92 & -47 31 2.1 & 15.01 & 0.02 & 13.44 & 0.02 & 12.36 & 0.02 & e \\
19 & 8 59 2.93 & -47 30 54.1 & 12.27 & 0.01 & 10.51 & 0.01 & 9.81 & 0.01 &  \\
20 & 8 59 2.93 & -47 30 54.1 & 12.14 & 0.02 & \nodata & \nodata & \nodata & \nodata &  \\
21 & 8 59 2.97 & -47 30 39.8 & 15.10 & 0.02 & 13.71 & 0.02 & 12.85 & 0.03 & e \\
22 & 8 59 2.98 & -47 30 51.7 & 17.88 & 0.10 & 15.90 & 0.06 & 14.12 & 0.05 & p \\
23 & 8 59 2.99 & -47 31 0.9 & \nodata & \nodata & 16.78 & 0.10 & 15.49 & 0.10 &  \\
24 & 8 59 3.00 & -47 30 44.4 & 16.25 & 0.03 & 13.54 & 0.02 & 11.00 & 0.01 & p \\
25 & 8 59 3.02 & -47 30 35.6 & 14.88 & 0.02 & 12.67 & 0.01 & 11.47 & 0.01 &  \\
\enddata
\tablecomments{In the Notes column, `e' refers to those sources with an
infrared excess, `p' to protostars, which are sources with an excess
greater than of that expected of a reddened T Tauri stars, and 'x' to
sources that have $J-H < 1$ and whose photometry is incorrect, for reasons
described in the text.  Units of right ascension are hours, minutes and
seconds, and unites of declination are degrees, arcminutes, and arcseconds.
Units of $J$, $J_{\rm err}$ $H$, $H_{\rm err}$ $K_s$ $(K_s)_{\rm err}$ are
magnitudes.  Table~\ref{src-list} is published in its entirety in the
electronic edition of the Astronomical Journal.   A portion is shown here
for guidance regarding its form and content.}
\label{src-list}
\end{deluxetable}

\clearpage
\begin{deluxetable}{c|cc}
\tablecolumns{3}
\tablewidth{0pt}
\tablecaption{Slopes for Larson plots for Taurus (Larson 1995) and the
clusters surveyed by Nakajima et al. (1998).  
$\gamma_S$ gives the power-law index at shorter angular separations
while $\gamma_L$ gives the power-law index at longer angular separations.}
\tablehead{\colhead{Region} & 
\colhead{$\gamma_S$} & \colhead{$\gamma_L$}}
\startdata
Orion OB & -1.6$\pm$0.4 & -0.15$\pm$0.02 \\
Orion A & \nodata & -0.23$\pm$0.02\\
Orion B & \nodata & -0.69$\pm$0.01\\
$\rho$ Oph & -2.5$\pm$0.3 & -0.36$\pm$0.06\\
Cha I & -2.1$\pm$0.2 & -0.57$\pm$0.04\\
Cha  & -2.4$\pm$0.5 & -0.55$\pm$0.03\\
Vela & \nodata & -0.61$\pm$0.02\\
Lupus &-2.2$\pm$0.4 & -0.82$\pm$0.13 \\
Taurus & -2.15 & -0.62\\
RCW~38 & \nodata & -1.32\\
\enddata
\label{larsontab}
\end{deluxetable}

%% FIGURES

\clearpage

\clearpage

\clearpage

\clearpage

\clearpage

\clearpage

\clearpage

\clearpage

\clearpage

\clearpage

\clearpage

\clearpage

\clearpage

\clearpage

\clearpage

\clearpage

\clearpage

\end{document}